# Analog Content-Addressable Memory from Complementary FeFETs

Xiwen Liu [a‡], Keshava Katti [a‡], Yunfei He [a], Paul Jacob [b], Claudia Richter [c], Uwe Schroeder [c], Santosh Kurinec [b], Pratik Chaudhari [a]*, Deep Jariwala [a]*.

[a] Electrical & Systems Engineering, University of Pennsylvania, Philadelphia, PA, USA.

[b] Electrical & Microelectronic Engineering, Rochester Institute of Technology, Rochester, NY, USA.

[c] Namlab/TU Dresden, Noethnitzer Strasse 64a 01187 Dresden, Germany.

*Corresponding authors: pratikac@seas.upenn.edu, dmj@seas.upenn.edu.

‡ Equal contribution.

**BIGGER PICTURE:** To address the increasing computational demands of artificial intelligence (AI), compute-in-memory (CIM) integrates memory and processing units into the same physical location, reducing the time and energy overhead of the system. AI applications of CIM residing beyond matrix multiplication, such as parallel search, remain relatively unexplored. We propose an analog content-addressable memory (ACAM) that executes parallel search in the analog domain using over 40 distinct match windows, alleviating the challenges of current parallel search architectures, including low density, high power consumption, and extensive peripheral components. From database querying to graph analytics, parallel search is indispensable for gleaning insights from massive, often unstructured, datasets. Hence, building efficient in-memory parallel search can markedly enhance the capabilities of existing CIM systems, opening avenues for more comprehensive data processing and facilitating richer AI applications.

**SUMMARY:** Despite recent advancements in non-volatile memory (NVM) for matrix multiplication, other critical data-intensive operations, like parallel search, remain largely overlooked. Current parallel search architectures, namely content-addressable memory (CAM), often use binary, which restrict density and functionality. We present an analog CAM (ACAM) cell, built on two complementary ferroelectric field-effect transistors (FeFETs), that performs parallel search in the analog domain with over 40 distinct match windows. ACAM not only offers a projected 3x denser memory architecture than ternary CAM (TCAM) but yields a 5% increase in inference accuracy on similarity search for few-shot learning simulated with the Omniglot dataset, with an estimated speedup per similarity search of more than 100x when compared to a central processing unit (CPU) and graphics processing unit (GPU) on scaled CMOS nodes. Furthermore, we demonstrate 1-step inference on a kernel regression model in ACAM, with simulation results indicating 1,000x faster inference than a CPU and GPU.

**INTRODUCTION**

In the rapidly evolving field of artificial intelligence (AI), processing vast amounts of data efficiently is critical. Computing architectures that incorporate in-memory processing are gaining traction, as they offer to potentially reduce power consumption and cost[1]. In particular, compute-in-memory (CIM) aims to reduce the energy-intensive data movement limiting traditional computing architectures by integrating memory and processing within the same physical location[2,3]. Recent studies have shown progress in employing non-volatile memory (NVM) devices to accelerate matrix multiplication within memory arrays, in turn aiding execution of complex AI tasks[4-7]. However, expanding these CIM architectures to accommodate other operations intrinsic to AI, notably parallel search, presents a considerable challenge[6-8]. We define parallel search to be a single-step operation of comparing a query vector to an array of $m$ stored vectors and returning a degree of match between the query and each stored vector.

Content-addressable memory (CAM) is well-suited to execute operations that match an input datum to a collection of data patterns stored in the CAM array[9]. If this operation occurs in parallel, then we can obtain high-throughput comparisons with minimal latency, a distinct advantage in applications such as network routing, associative computing, and database management systems[10-12]. In standard silicon-based complementary metal-oxide semiconductor (CMOS) architectures, the construction of a single CAM cell requires approximately 16 transistors, typically configured with static random-access memory (SRAM), leading to significant power consumption and relatively low memory density[13]. NVM has emerged as a promising substitute within CAM, given its reconfigurable functionality, as well as its superior area and energy efficiency. There have already been successful demonstrations of ternary CAM (TCAM) implemented using various NVM technologies[14-18]. However, most CAM designs based on NVM employ designs akin to traditional SRAM, where the memory device is restricted to encoding binary states exclusively and is used for identifying exact matches or mismatches. Given that CAM designs are primarily in the binary domain, current architectures frequently require extensive analog-to-digital conversion (ADC) and digital-to-analog conversion (DAC). ADC and DAC serve as pivotal links between digital memory and analog components (e.g., sensors and actuators), translating various inputs into a form that CAM can handle. The consistent need for conversion significantly diminishes both power and speed, a particularly non-ideal energy overhead in the context of edge computing[19]. More recently, analog CAM (ACAM), which leverages programmable analog conductances within NVM, has been proposed[20-22]. As of now, the hardware development for ACAM poses substantial difficulties and has hence rarely been reported. A recently proposed ACAM cell based on resistive RAM (ReRAM) demands 6 silicon CMOS transistors, resulting in low area efficiency, high parasitic components, and elevated static power consumption[21]. Another proposed ACAM with ferroelectric field-effect transistors (FeFETs) is a theoretical design that requires two search lines and a complex analog converter circuit to perform the search signal linear reversion function[22].

In this work, we report an experimental demonstration of a non-volatile and compact ACAM cell based on two complementary FeFETs. Our ACAM cell utilizes the adjustable threshold voltages of FeFET devices to store over 40 distinct windows within each cell and examines an analog input against this stored range to identify a match or a mismatch. This cell neither requires additional transistors nor an external analog converter circuit — it only requires a single search line. Two key advantages arise from this design: (i) Parallel search is conducted where the data resides, eliminating the power and latency costs of data transportation between separate computing and memory units; (ii) Computation and analog-to-digital conversion are merged, offering higher memory bit-densities and reduced power usage for a broader set of applications.

We go on to evaluate the performance of our ACAM design for similarity search in a few-shot learning application, using matching networks for inference on the Omniglot dataset and benchmarking both the accuracy and runtime of ACAM against a central processing unit (CPU), graphics processing unit (GPU), and TCAM. Additionally, we can interpret the match operation in ACAM as a kernel and hence deploy ACAM for kernel regression, a model for fitting non-linear functions. Kernels in machine learning are used to model the similarity between two inputs and form the basis for a wide variety of models, including support vector machines (SVMs), kernel regression and classification models, and, most recently, neural tangent kernels for deep learning. Such models typically require intensive computation of pairwise similarity between a test data point and each training data point during inference that could benefit greatly from parallel search.

## RESULTS
### ACAM Cell Design and Measurement

Different from conventional CAM architectures in which the digital query and keys are compared for identifying exact matches or mismatches (shown in Fig. 1a), Fig. 1b-c illustrate the concept and design of the proposed ACAM, where analog voltage values are supplied as the query to the ACAM and compared against the analog ranges encoded by adjustable threshold voltages of FeFET devices. In this work, the cell structure of ACAM can be significantly simplified by using just two complementary FeFETs (as shown in Fig. 1c), in which the n-channel metal-oxide semiconductor (NMOS) FeFET is colored in red, and the p-channel metal-oxide semiconductor (PMOS) FeFET is colored in blue. The analog input search data get converted into voltage amplitudes which are applied along the search line (SL), while the stored analog range is set by the programmed threshold voltages of the cell's two complementary FeFETs. The search operation begins by pre-charging each match line (ML) to a high voltage level. The ML remains high (indicating a match) only when the discharge currents at all ACAM channels are minimal. That is, all attached CAM cells of a row match to the input data. Otherwise, the significant discharge currents of any ACAM channels result in a voltage drop (signifying a mismatch) on the ML. FeFETs work by utilizing positive or negative gate pulses to direct the ferroelectric polarization toward the channel or gate metal, setting the FeFET to a low or high threshold voltage state, respectively. As opposed to other NVM devices that demand substantial DC conduction current for memory write, FeFETs excel in write energy efficiency, as they solely rely on the electric field to switch the polarization. The devices used in this work are FeFETs with hafnium zirconium oxide (HZO) as the ferroelectric gate dielectric integrated on traditional silicon (Si) CMOS transistor technology (see methods for more details)[23]. As a result, these devices can be generalized to any complementary FeFET technology.

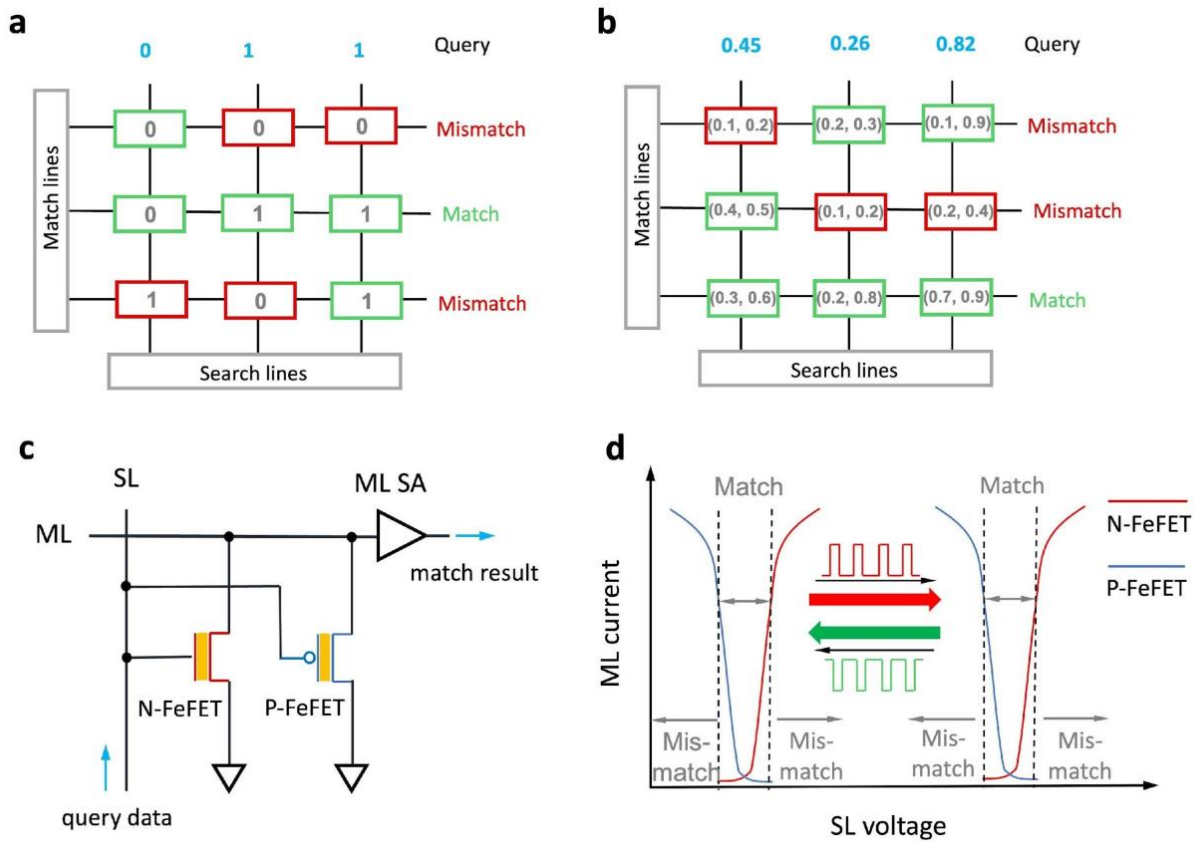

**Fig. 1. ACAM concept and design.** (**a**) Schematic diagram of digital CAM architectures in which digital query and keys are compared for identifying exact matches or mismatches. Considering that current CAM designs primarily operate in the digital domain, their bit-density and functionality are limited. These architectures often rely heavily on ADC operations to serve as critical interfaces between digital memory and analog components, which significantly reduces both power and speed, a considerable energy overhead, especially in the context of edge computing. (**b**) Schematic diagram of analog CAM architecture where analog query is compared with analog intervals, aiming to identify whether it falls within (match) or outside of these intervals (mismatch). (**c**) We substantially simplify ACAM cell structure by using just two complementary FeFETs, where the NMOS FeFET is represented in red and PMOS FeFET in blue. Analog input search data are converted into voltage amplitudes and applied along the search line (SL), while stored analog range is determined by the programmed threshold voltages of the two complementary FeFETs within the cell. Similarity between query and keys outputted on match

line (ML). (**d**) Upper and lower limit of each match window can be programmed by the threshold voltages of NMOS FeFET and PMOS FeFET, respectively. Furthermore, threshold voltage of both NMOS FeFETs and PMOS FeFETs can be adjusted using electrical pulses. This offers an additional level of adaptability to the ACAM cell by providing precise control over FeFETs' properties, thereby enabling the storage of continuous intervals as a match window (for more details, refer to the supplementary materials).

We will now discuss how the complementary FeFETs in the ACAM cell store analog match windows to compare against the input search value. To begin, the ML is connected to NMOS FeFET and PMOS FeFET in parallel. Further, the ML stays high for a match outcome when the SL voltage is lower than the threshold voltage of the NMOS FeFET and higher than the threshold voltage of the PMOS FeFET, thus maintaining both FeFET channels in a cut-off state. As shown in Fig. 1d, the upper and lower bound of each match window can be programmed by the threshold voltages of the NMOS FeFET and PMOS FeFET, respectively. When the SL voltage ($V_{SL}$) exceeds the threshold voltage of the NMOS FeFET, it gets turned on. This leads to a substantial discharge current between the ML and ground (GND), primarily through the NMOS FeFET, which in turn causes a voltage drop in the ML that produces a mismatch outcome. On the other hand, when $V_{SL}$ is below the threshold voltage of the PMOS FeFET, it is turned on, leading to a substantial discharge current between the ML and GND, predominantly across the PMOS FeFET. This results in a voltage drop in the ML, also producing a mismatch outcome. When the SL voltage resides within the threshold voltages of both NMOS FeFETs and PMOS FeFETs, both types of FeFETs are in a cut-off state. Consequently, no discharging current will traverse through either of the FeFETs between the ML and GND, and the ML remains at a high level, denoting a match result. In addition, the threshold voltage of both NMOS FeFETs and PMOS FeFETs can be programmed using electrical pulses, as shown in Fig. 1d. This allows for precise and efficient tuning of the FeFETs' characteristics, adding an extra layer of flexibility to the ACAM cell and enabling it to store continuous intervals as a match window.

In order to substantiate our circuit design and delve deeper into the concept of complementary FeFET-based ACAM, we perform experimental measurements of ACAM circuit operation on HZO Si CMOS FeFET chips (for more details, refer to the supplementary materials). We first present the programmable threshold voltages in both NMOS FeFETs and PMOS FeFETs. These threshold voltages serve to define the upper and lower bounds of the ACAM match window, respectively. We note that we achieve these programmable threshold voltages through partial polarization switching, a process that can be implemented by biasing short electrical pulses at the gate electrode of the FeFETs[24]. Fig. 2a illustrates the gradual programming of the $I_D$-$V_G$ transfer curve in the NMOS FeFET using a series of stepwise gate voltage pulse modulations, ranging from 3 – 4 V. This gradual programming enables the NMOS FeFET to present 10 unique transfer curves, each exhibiting a high level of linearity. Fig. 2b demonstrates that, akin to the NMOS FeFET, the PMOS FeFET is also capable of undergoing analog threshold voltage programming induced by electrical pulses. Positive pulses will make the threshold voltage more positive for both NMOS FeFETs and PMOS FeFETs, as the threshold voltage depends on the absolute value of the amount of polarization in the ferroelectric dielectric. It is noteworthy that the application of a positive pulse in both NMOS FeFETs and PMOS FeFETs elevates their threshold voltages. This phenomenon occurs because these threshold voltages hinge upon the absolute quantity of polarization in the ferroelectric dielectric in both device types.

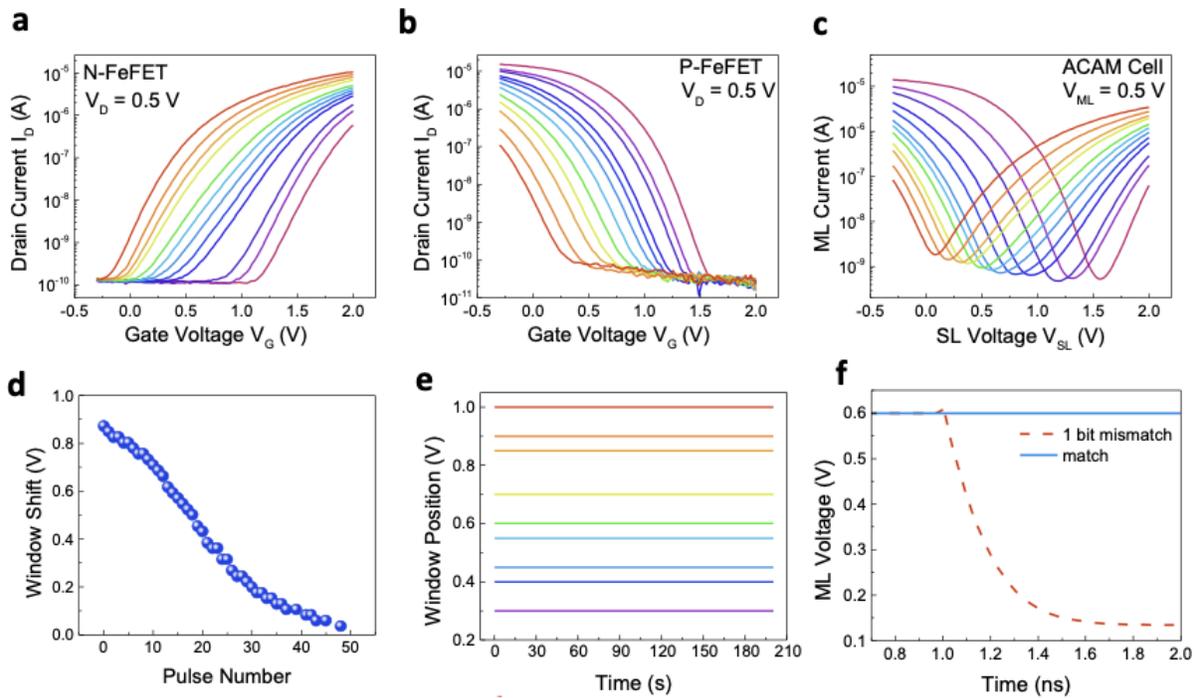

**Fig. 2. Experimental demonstration of ACAM.** (**a**) Analog programming of the $I_D$-$V_G$ transfer curve in NMOS FeFET via stepwise gate voltage pulses of ~1 μs, ranging between 3 – 4 V. NMOS FeFET successively yields 10 distinct transfer curves due to this graduated programming. (**b**) PMOS FeFET, like its NMOS FeFET counterpart, is also shown to be capable of analog threshold voltage programming through electrical pulses. It is important to note that threshold voltages of both NMOS FeFETs and PMOS FeFETs are influenced in the same manner by applied pulses, as these voltages are contingent on the absolute value of polarization within the ferroelectric dielectric of each device. (**c**) Current passing through ML is measured when sweeping voltage is applied on SL. By applying different voltage pulses, the central position of the match window can be adjusted, hence generating multiple match windows at various positions. (**d**) Position of ACAM match window can be analogously adjusted, thus providing over 40 distinct match windows. (**e**) Demonstration of 9 stored ranges exhibiting stable retention without significant degradation indicates that ACAM is non-volatile. (**f**) Simulated ML discharge behavior during search operation, specifically for all-match and 1-bit-mismatch states, in a 64-row, 64-column ACAM array on 45 nm Si CMOS technology. Simulations indicate notable distinction, with estimated ML delay of 0.136 ns.

Subsequently, we perform a proof-of-concept measurement of the performance of the designed ACAM cell. We first define a match window by programming the threshold voltages of the NMOS FeFETs and PMOS FeFETs via electrical pulses over the gate terminal. Then, we measured the ML discharge current with a sweeping SL voltage. As shown in Fig. 2c, for each fixed match window, the ML current remains low near the center (signifying a match) and drastically surges as the SL voltage deviates from the window center (signifying a mismatch). Moreover, we observe that by applying voltage pulses, we are able to vary the central position of the match window, thereby creating multiple match windows at different positions. As depicted in Fig. 2d, not only is the input in the analog domain, but the location of the ACAM match window can also be analogously programmed, offering more than 40 distinct match windows. This demonstrates the successful operation of the FeFETs in an ACAM cell. As illustrated in Fig. 2e, stable retention without noticeable degradation is exhibited for 9 distinct stored ranges, which suggests that ACAM is not only non-volatile but does not require static power for range retention or require frequent refreshing once programmed. For the fast data operations discussed in this paper, a retention period of hundreds of seconds is more than sufficient. In order to evaluate latency of the search operation in ACAM, we carry out extensive circuit-level analysis of a 64-row, 64-column ACAM array, built upon 45 nm Si CMOS technology and inclusive of peripheral circuits (for more details, refer to the supplementary materials). Fig. 2f presents the simulated ML discharge patterns during a worst-case search operation (considering all-match and 1-bit-mismatch states), clearly demonstrating a significant difference, with the delay in the ML quantified as 0.136 ns.

**ACAM for Similarity Search**

An ACAM cell can store real-valued intervals, as opposed to bits in a binary or ternary memory cell, giving us a new way to search and retrieve directly in the analog domain without converting signals into their digital counterparts. Similarity search is a key problem in machine learning, forming the building block for many applications, such as content retrieval, where a user seeks to recover sentences, audio files, or images that are similar to a given query. Machine learning models, including *k*-nearest neighbor classifiers, support vector machines (SVMs), and kernel machines, have similarity search as a key component of their inference mechanism. A growing body of existing work has therefore been devoted to accelerating these computations[11,25-27]. We show that ACAM cells can be used to store data in its native, real-valued format. Furthermore, compact arrays of such cells can perform similarity search for a user query at a very low latency, which we will demonstrate for inference in few-shot learning.

As image classification systems begin to tackle more and more classes, the cost of annotating a massive number of images, as well as the difficulty of procuring images of rare categories, increases. This challenge has fueled interest in few-shot learning, a type of learning where only a few labeled samples per class are available for training. The key idea behind current few-shot learning methods is to train a model to distinguish inputs, say images of different categories, from each other. This strategy is different from classical supervised learning, which seeks to predict the category of an input image. Features learned in such models using large datasets, such as the ImageNet-21K dataset (14.2 million images from 21,814 different categories), can be fruitfully used to distinguish between images of entirely new categories using very few new images (i.e., "few-shot")[28]. One of the key steps in doing so involves calculating the similarities between the features of few-shot labeled data and the test data, as suggested by an algorithm called "matching networks."[29]. For $k$-shot learning across $n$ different classes (typically, $k$ ranges from 1 to 10 and $n$ can range anywhere from 5 to 100), the centroid of the features of the $k$ images of each class is computed. Given a test image (also called the "query"), the similarity (e.g., Hamming distance, inner product) of the features of the test image, which could be from any of the $n$ classes, is computed to find the centroid closest to it. See the schematic in Fig. 3a for an example where the unknown query sample (i.e., a Dachshund breed) should be matched to the Yellow Labrador breed, as they are both dogs, in the 4-way, 1-shot support set containing a dog, cat, pig, and fish.

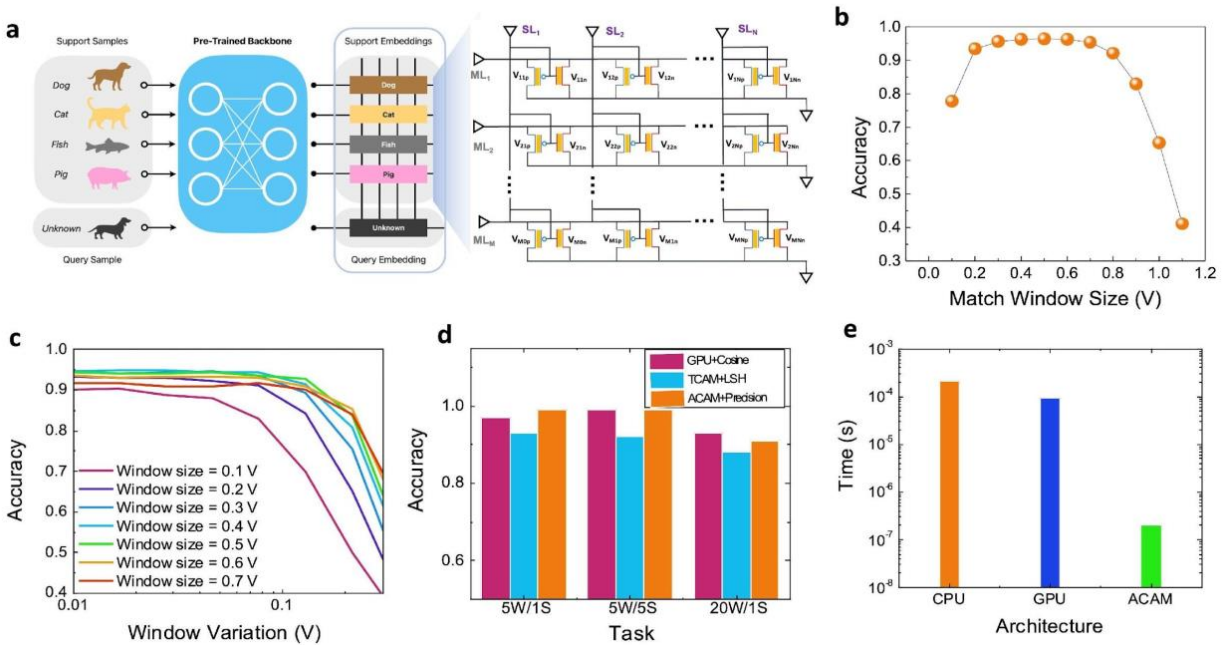

**Fig. 3. Matching networks inference for few-shot learning on ACAM.** (**a**) 4-way, 1-shot few-shot learning episode with dog, cat, fish, and pig encoded as 64-dimension support embedding by pre-trained backbone. Each element of support embedding stored in ACAM cell, while query

embedding placed on search line. Match line outputs 4-dimensional vector scoring similarity between query embedding and 4 support embeddings, with query label given as label of support sample with largest similarity score. (**b** – **e**) Omniglot dataset partitioned into *n*-way, *k*-shot episodes. A 4-layer convolutional neural network (CNN), called the backbone, outputs 64-dimensional embedding for each sample. Each layer in backbone has 3×3 kernel with 64 filters, batch normalization, ReLU non-linearity, and a 2×2 max-pooling. Backbone trained on 1,200 characters, while inference occurs on remaining 423 characters. (**b**) Optimal match window size shown to be 0.4 V in order to maximize inference accuracy. (**c**) Inference accuracy remains approximately constant at 90 – 95% for match window size ranging between 0.2 – 0.7 V for window variation up to 0.1 V. (**d**) Benchmark of inference accuracy shows ACAM outperforms TCAM for 5-way and 20-way tasks and remains comparable to GPU. TCAM+LSH data sourced from[11] which uses a 28 nm CMOS node. But ACAM is a notably denser memory architecture, requiring only 64 cells per 64-dimensional embedding, while TCAM needs on the order of 128 or 256 cells. (**e**) Amount of time in seconds to perform inference on ACAM (45 nm node) for one query sample shown to be more than 2 orders of magnitude less than both CPU (14 nm node) and GPU (12 nm node).

Similarity search in few-shot learning can be mapped to ACAM as follows. In software, the support embeddings are stored in an array of size $n \cdot k \times 64$, while the query sample is represented as a 64-dimensional array. Similarly, an $(n \cdot k \times 64)$-dimensional ACAM array is used to store the features of few-shot labeled images embeddings, where the $(i,j)$th element corresponds to the $j$th element of $i$th support embedding. Programming of the support embeddings into ACAM involves selecting a window size, typically held constant across all ACAM cells and then positioning the window such that the $(i,j)$th element lies at the center of its window for all $i = 1, \ldots, n \cdot k$ and $j = 1, \ldots, 64$. Recall that the central position of the match window can be adjusted by gradually applying voltage pulses to the NMOS FeFET or PMOS FeFET, pictured in each ACAM cell of the Fig. 3a circuit diagram. On the other hand, the query embedding is programmed onto the search lines, shown at the top of each column in the Fig. 3a circuit diagram, such that the $j$th element of the query embedding lies on the $j$th SL of the $n \cdot k \times 64$ ACAM array for all $j = 1, \ldots, 64$. Binary and ternary CAM compute conventional Hamming distance, which checks for exact match between bits and can be written as $S_{\text{digital}}(\mathbf{s}_i, \mathbf{q}) = \sum_{j=1}^{64} \mathbf{1}(\mathbf{s}_i[j] = \mathbf{q}[j])$, where $\mathbf{1}(\cdot)$ is the indicator function. In contrast, ACAM computes a "generalized" Hamming distance within the analog domain, which can be expressed as $S_{\text{analog}}(\mathbf{s}_i, \mathbf{q}) = \sum_{j=1}^{64} \mathbf{1}(\mathbf{q}[j] \in [a_{i,j}, b_{i,j}])$, where $a_{i,j}$ and $b_{i,j}$ are the endpoints of the match window in the ACAM cell corresponding to the $j$th element of $i$th support embedding. To reiterate, the main advantage of ACAM over binary and ternary CAM is that it computes similarity directly in the analog domain via generalized Hamming distance.

Selection of the matching window size and impact of added noise within ACAM will now be discussed in the context of similarity search. Figs. 3b-c are benchmarked with a 5-way, 5-shot inference task on the Omniglot dataset, which contains 1,623 characters from 50 different alphabets and is down-sampled such that each image is 28×28. Further, all support embeddings stored in ACAM are quantized to lie within the range $[-0.3, 2.0]$ V to align with the range of operation shown in Figs. 2a-c. When programming the support embeddings into ACAM, the first step is to select a window size. A match window range of within $[0.0, 1.0]$ V is sufficiently expressive to achieve the highest inference accuracy, which occurs at a match window size of 0.4 V as seen in Fig. 3b. While it is not possible to avoid noise within CAM, it is nonetheless possible to show that ACAM is robust to any perturbations in its match window when used for similarity search in few-shot learning. Fig. 3c shows the result of noise sampled from a normal distribution with mean $\mu = 0$ and standard deviation $\sigma$ being added to the match window, illustrating that a window variation of up to $\pm 0.1$ V, which is $\pm 4.35\%$ of the entire range of ACAM operation, is tolerable without incurring a decrease in inference accuracy of more than $10\%$.

It is important to compare the density of ACAM versus TCAM. Analog representations in ACAM are far more expressive than the digital representations in TCAM, which in turn yields denser memory storage while maintaining high-accuracy results. Specifically, binary/ternary representations require the use of locality-sensitive binary codes. It is shown that binary classification on 14,871 images of dimension 320 taken from the LabelMe database result in a diverse set of precision-recall curves when locality-sensitive binary codes are used to embed the test data with various code sizes[30]. Precision represents the fraction of positive predictions that actually belong to the positive class, while recall is the fraction of positive predictions out of all positive instances in the dataset. The goal is to maximize both. For a given recall value of 0.8, the precision varies between 0.4, 0.65, and 0.8 for 256-, 512-, and 1,024-bit codes, respectively. It is observed that larger bit codes (e.g., 512 or 1,024) are more suitable for real-world applications. Within matching networks, the backbone embeds 784-dimension Omniglot data into 64 dimensions. If 512- or 1,024-bit codes are strongly suggested for 320-dimensional data, it is reasonable to expect at least 128- or 256-bit codes for the Omniglot dataset, which translates to an $n \times 128$ or $n \times 256$ TCAM array. In sharp contrast, ACAM requires only an $n \times 64$ array, meaning ACAM is a 3x denser memory architecture than TCAM.

Lastly, the performance of matching networks with ACAM will be benchmarked with respect to accuracy and runtime. After executing inference on 423 characters, Fig. 3d shows the accuracy results of ACAM (45 nm node) are on average 5% higher than those of TCAM (28 nm node) and comparable to those of the GPU (NVIDIA Tesla T4, 12 nm node), where TCAM+LSH data sourced from[11]. Additionally, the time incurred for a single inference operation in ACAM is significantly less than both CPU (Intel Xeon Platinum, 14 nm node) and GPU by more than 2 orders of magnitude as shown in Fig. 3e. To summarize, ACAM represents an efficient alternative to CPU, GPU, and TCAM for similarity search in few-shot learning that not only outperforms TCAM in its inference results (while remaining comparable to the GPU) but does so with a denser memory architecture that enables faster, parallel computation. It is further worth noting that the above is a conservative estimate since the ACAM is computed on a 45 nm Si CMOS mode, with Si

FeFETs currently available in the more advanced 28 nm technology node, while the CPU and GPU results follows from 14 nm and 12 nm nodes, respectfully[11,23]. Further details on the software simulation can be found in the supplementary materials.

**ACAM for Kernel Regression**

We next demonstrate how to use ACAM for inference in a kernel regression machine[31]. Given two inputs $\mathbf{x}$ and $\mathbf{x}'$, which are $d$-dimensional vectors, a kernel is a function $K(\mathbf{x}, \mathbf{x}')$ that can be understood as an estimate of the similarity between the two inputs. Such a function can be used to make predictions as follows. Given a training dataset $\mathcal{D}_{\text{Tr}} = \{(\mathbf{x}_i, y_i)\}_{i=1}^m$, where $\mathbf{x}_i$ are the inputs and $y_i \in \mathbb{R}$ are the outputs, we compute the "Gram matrix" $\mathbf{K}[i,j] = K(\mathbf{x}_i, \mathbf{x}_j)$ whose entries are the pairwise similarities between inputs in the training set. Predictions on a new test datum $\mathbf{x}$ are computed as $\hat{y}(\mathbf{x}) = \sum_{i=1}^m \hat{\alpha}_i K(\mathbf{x}_i, \mathbf{x})$, where $\hat{\boldsymbol{\alpha}} = (\mathbf{K} + \lambda m \mathbf{I}_m)^{-1} \mathbf{y}$ is the vector of coefficients that are used to weigh the true targets of the $m$ samples, denoted by $\mathbf{y}$, to make the predictions on the test datum $\mathbf{x}$. Targets of input samples that are more similar to the test datum are upweighted in the above summation. Note that the parameter $\lambda$ is known as the ridge regression constant, allowing us to regularize the fitting procedure in situations when there are few samples in the training dataset. Observe that it biases the diagonal of the Gram matrix $\mathbf{K}$ and effectively makes each input sample play a role in making predictions on the test datum. Mathematical details of this procedure are provided in the supplementary materials.

There are many kernels $K(\mathbf{x}, \mathbf{x}')$ that can be used in this procedure. One popular choice involves a radial basis function (RBF) kernel $K(\mathbf{x}, \mathbf{x}') = \exp(-\|\mathbf{x} - \mathbf{x}'\|_2^2 / 2\gamma^2)$, which computes the probability of the test datum being drawn from a Gaussian distribution centered at one of the training data points, as seen in Fig. 4a(i). Another version consists of the Laplace kernel $K(\mathbf{x}, \mathbf{x}') = \exp(-c\|\mathbf{x} - \mathbf{x}'\|)$. Constants like $\gamma$ and $c$ in these expressions are considered hyperparameters and are chosen using cross-validation. Computing the prediction in a kernel regression machine involves computing the summation above, and even if coefficients $\hat{\boldsymbol{\alpha}}$ are computed beforehand, it is necessary to compute the $m$ terms $K(\mathbf{x}_i, \mathbf{x})$ for each new test datum $\mathbf{x}$. Further, all the $m$ training samples must be stored in memory.

ACAM offers a different way to implement kernel regression. Observe that the current-voltage characteristic curve in an ACAM computes a kernel. The transfer curves shown in Fig. 2c can be approximately described by the expression $\exp(|V_G - \mu|^2 / 2\gamma^2)$, where $V_G$ is the applied gate voltage, and $\mu$ is the mean of the match window. This expression can be equivalently written using the previous kernel notation as $\exp(-|x - x'|^2 / 2\gamma^2)$. Since this expression looks like a parabola, a "surrogate" Gaussian kernel is achieved by the following constant-time negation, summation, and maximization operations, yielding $K^{\text{ACAM}}(\mathbf{x}, \mathbf{x}') = \max\{0, 2 - \exp(|\mathbf{x} - \mathbf{x}'|^2 / 2\gamma^2)\}$. See Supplementary Fig. 4a for an overlaid plot of the Gaussian kernel and surrogate Gaussian kernel. This surrogate kernel is what ACAM outputs on the match line for some stored vector $\mathbf{x}$ and query vector $\mathbf{x}'$.

It will be shown that ACAM enables rapid, 1-step inference. All that is required is for a test data point to be placed on the search lines of the ACAM. The surrogate Gaussian kernel can be computed in $O(1)$ time. Fig. 4a(ii) shows how the multiplication with $\hat{\boldsymbol{\alpha}}$ can also be designed to occur within the same step by placing $\hat{\alpha}_i$ on the drain of both the NMOS FeFET and PMOS FeFET in the $i$th ACAM cell for all $i = 1, \ldots, m$, which linearly scales each output $K(\mathbf{x}_i, \mathbf{x}')$ by $\hat{\alpha}_i$. Since the ML sums input currents, the ML output becomes $\sum_{i=1}^{m} \hat{\alpha}_i K(\mathbf{x}_i, \mathbf{x}')$, the predicted label $\hat{f}(\mathbf{x}')$ under kernel regression. Hence, ACAM is able to achieve 1-step inference for kernel regression.

To evaluate the ability of ACAM to perform such inference, 1-dimensional synthetic data was generated by randomly sampling $f(x) = \sin(5x)$, known as the ground-truth function, and adding noise randomly sampled from a normal distribution with mean $\mu = 0$ and standard deviation $\sigma = 0.2$. This process is then used to generate both the training and test data. Like in the case of few-shot learning, the ACAM-based kernel regression model exhibits robustness to noise. As seen in Fig. 4b, even with a window variation of up to 0.3 V, the mean squared error (MSE) of the fitted function only changes on the order of $10^{-3}$ to $10^{-2}$. Furthermore, Fig. 4b shows that the reduction in MSE saturates at 4 bits, after which additional bits do not lead to diminished inference error. Since ACAM operates at about 4 bits per cell, Fig. 4b suggests that neither quantization nor noise significantly degrade the performance of ACAM-based inference. Figs. 4c(i)-(iii) illustrate the predictions (red) made by ACAM on the scattered test data points (blue). The quality of the fit can be observed by the closeness between the ground-truth function (black) and the ACAM predictions. Note that the surrogate Gaussian kernel has a parameter $\gamma$ that determines the width of the function and, by extension, the slope of the current-voltage plot. From Fig. 2c, this value can be approximated as $\gamma = 0.1$ V. Figs. 4c(i)-(iii) show the test predictions for $\gamma > 0.1$ V, $\gamma = 0.1$ V, and $\gamma < 0.1$ V, corresponding to underfitting, optimal, and overfitting conditions, respectively, and Supplementary Figs. 4b-c show how training and test data can be scaled in order to ensure that $\gamma = 0.1$ V remains the optimal condition. To further evaluate the impact of noise, we determine the predictions outputted by ACAM under noisy conditions $\mathbf{y}_{\text{noise}}$. After that, we compute the residuals $\mathbf{y}_{\text{noise}} - \mathbf{y}_{\text{ground-truth}}$, which quantify how close the predicted labels are to the ground-truth labels and plot them as a histogram in Figs. 4c(iv)-(vi). As expected, the histogram traces out a Gaussian distribution with mean $\mu = 0$ for a window variation of 0.01 V, meaning the vast majority of predicted labels were the same as the ground-truth labels (shown in Fig. 4c(iv)). For a window variation of 0.08 V, it can be seen that the fitted Gaussian is no longer quite centered at mean $\mu = 0$, while the variance has increased (shown in Fig. 4c(v)). The trend in the means and variances of Gaussian functions fitted to the residuals as noise increases can be seen in Fig. 4c(vi). Though the variances increase quadratically, the means remain close to $\mu = 0$, which further confirms that ACAM remains tolerant to noise in inference.

Lastly, benchmarking of the time required to complete inference on a single test data point shows that ACAM (45 nm node), performs 3 orders of magnitude faster than CPU and GPU (12 nm nodes) similar to the above demonstration of few-shot learning. See supplementary materials for a justification on why CPU and GPU have been placed in the same column. Furthermore, Fig. 4d shows that both CPU and GPU perform on the order of $64 \cdot 64 = 4{,}096$ floating-point operations (FLOPs) during inference, while ACAM requires only 1. To recapitulate, this section has presented a 1-step inference procedure for kernel regression that leverages the strengths of ACAM in computing a (surrogate) Gaussian kernel. This result extends previous work of CIM hardware for linear regression to non-linear regression and opens the door to future applications of ACAM in other attention- or kernel-based learning[32].

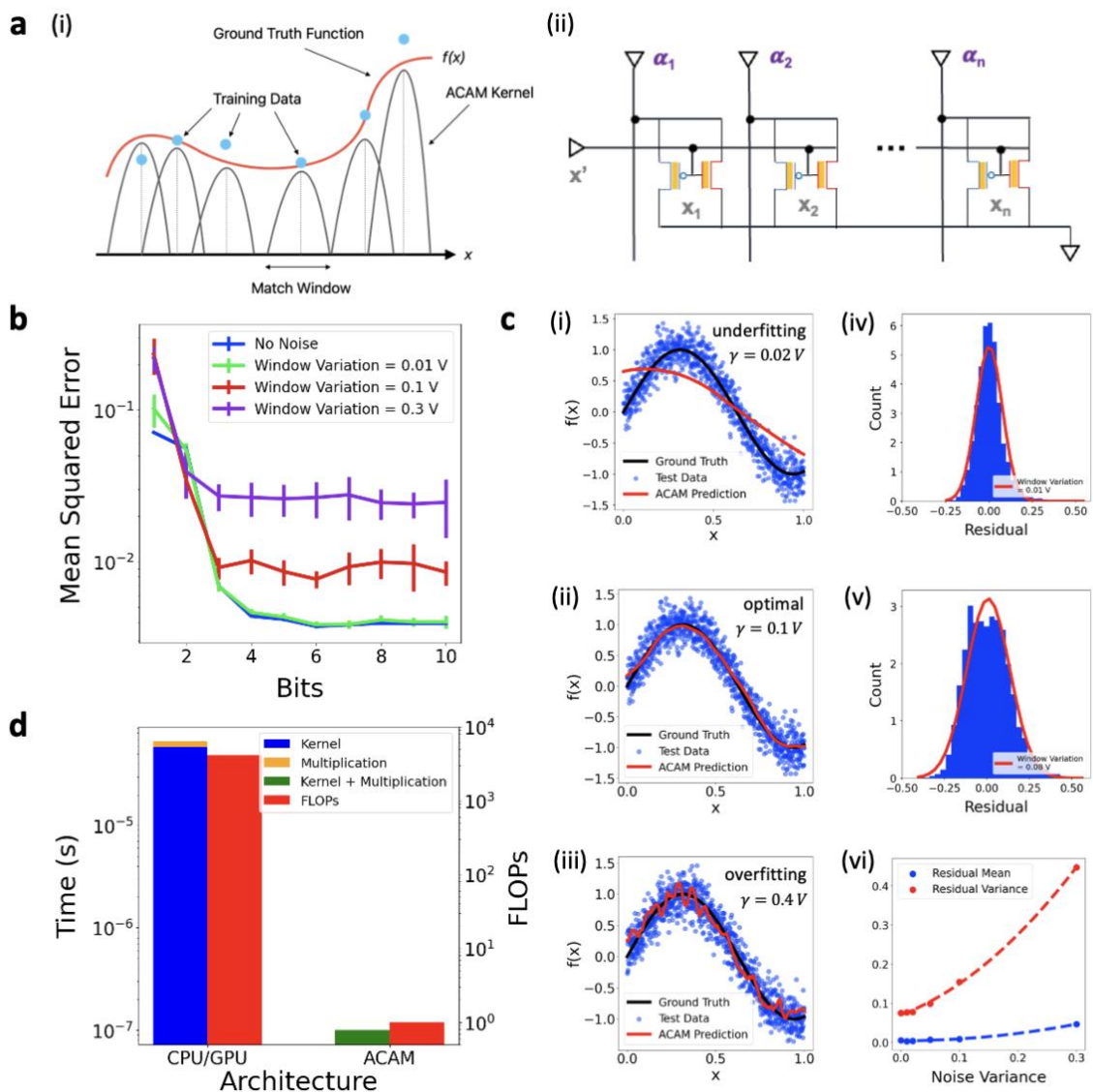

**Fig. 4. Inference for kernel regression model on ACAM.** (**a**) (*i*) Fitting kernel regression model can be thought of as summing a set of Gaussian functions, with mean centered at each data point, to get a function that reflects all the given data. (*ii*) Circuit that computes predicted label $\hat{y}(\mathbf{x}) =$

$\sum_{i=1}^{m} \hat{\alpha}_i K^{\text{ACAM}}(\mathbf{x}_i, \mathbf{x})$ during inference phase in 1 step. (**b**) Mean squared error (MSE) remains less than 0.03 for 4-bit quantization, approximately what ACAM operates at, with window variation up to 0.3 V. (**c**) (*i*) Underfitting exhibited by ACAM predictions (red) on test data (blue) with too large of a kernel parameter $\gamma = 0.4$ V. (*ii*) Optimal fitting exhibited by ACAM predictions (red) on test data (blue) with just right of a kernel parameter $\gamma = 0.1$ V, the typical width of ACAM surrogate Gaussian kernel. (*iii*) Overfitting exhibited by ACAM predictions (red) on test data (blue) with too small of a kernel parameter $\gamma = 0.02$ V. (*iv*) Histogram of residual $\mathbf{y}_{\text{noise}} - \mathbf{y}_{\text{ground-truth}}$ for small amount of Gaussian noise ($\mu = 0, \sigma = 0.01$) added to ACAM cells. (*v*) Histogram of residual $\mathbf{y}_{\text{noise}} - \mathbf{y}_{\text{ground-truth}}$ for moderate amount of Gaussian noise ($\mu = 0, \sigma = 0.08$) added to ACAM cells. (*vi*) Trend of means and variances of Gaussian fit to residual histogram as noise increases, showing that ACAM boasts robustness to noise, as residual means remains close to $\mu = 0$. (**d**) Benchmark of CPU and GPU (12 nm node) versus ACAM (45 nm node) in terms of time in seconds and number of floating-point operations (FLOPs) needed to perform inference on a single test datum, with ACAM again outperforming its counterparts by about 3 orders of magnitude.

**Conclusion**

In summary, we demonstrate an ACAM CIM architecture based on complementary FeFETs. We experimentally validate the operation and function of individual ACAM cells on HZO Si complementary FeFET devices and then use simulation tools to compare ACAM performance in similarity search and kernel regression. We demonstrate that our ACAM architecture at 45 nm CMOS node can outperform CPU and GPU at 12 nm CMOS node in similarity search by two orders of magnitude and by three in kernel regression. Given these advantages of our ACAM CIM over CPU and GPU, the question becomes what kind of CIM-based pattern matching architecture to use. Among CAMs, the choice is between digital TCAM or our presented ACAM. TCAM requires extensive ADC and DAC operations that significantly reduce both power and speed, providing a limited advantage in pattern matching performance over algorithms on CPU and GPU. As a result, the proposed ACAM in this work represents a strong candidate for accelerating pattern matching in machine learning. In the future, CAM-based pattern matching architectures could be mapped to other machine learning or robotics tasks, including visual scene understanding, kernel support vector machines, or even attention in large-language models, like transformers.

**EXPERIMENTAL PROCEDURES**
**Resource Availability**
*Lead Contact*
Further information and requests for resources should be directed to the lead contact, Deep Jariwala (dmj@seas.upenn.edu).

*Materials Availability*
This study did not generate new unique reagents.

*Data and Code Availability*

The data and code that support the conclusions of this study are also available from the lead contacts upon reasonable request.

**Device Fabrication**

NMOS FeFETs and PMOS FeFETs were fabricated in the Rochester Institute of Technology (RIT) student-run fabrication facility on $1-10\,\Omega\cdot\text{cm}$ base resistivity silicon wafers using the RIT CMOS process consisting of LOCOS isolated field-effect transistors with ion-implanted source and drain regions. Custom masks were designed using Mentor Graphics Pyxis and fabricated using the Heidelberg DWL 66+ laser writer. Photolithography was performed using an i-line ASML PAS 5500/200 stepper. FeFET device fabrication was performed at Namlab, Germany, with 11 nm thick $\text{Hf}_{0.5}\text{Zr}_{0.5}\text{O}_2$ (HZO) films deposited via atomic layer deposition (ALD) by alternating cycles of $\text{HfO}_2$ and $\text{ZrO}_2$ using $\text{HfCp(NMe}_2)_3$ and $\text{ZrCp(NMe}_2)_3$ as metal-organic precursors and ozone as an oxidant on a native $\text{SiO}_2$ layer on Si. A TiN top electrode was deposited via sputtering under ultra-high vacuum. Both films were annealed at 500°C for 20 s in $\text{N}_2$. Measurements on a metal-ferroelectric-metal capacitor structure fabricated in the same deposition run as the FeFET devices showed remanent polarization ($\text{P}_r$) values of about 20 µC/cm².

**Device Characterization and Simulation**

Current-voltage measurements were performed in air at ambient temperature using a Keithley 4200A semiconductor characterization system. The ACAM array inference is simulated using SPICE simulation. This simulation fully incorporates the ACAM array, under the assumption that the peripheral circuits, inclusive of the MLSA, are grounded in 45 nm silicon CMOS technology. In terms of inference applications, the focus is solely on benchmarking the search operation. Given that inference does not include a write operation, the characteristics of the ACAM cell are simulated using parameters from 45 nm silicon CMOS technology transistors.

**ACKNOWLEDGEMENTS**

K.K. acknowledges support from the National Science Foundation (NSF) Graduate Research Fellowship Program (GRFP), Fellow ID: 2022338725. D.J. and X.L. acknowledge partial support from Intel Rising Star Faculty Award. P.C. was supported by grants from the National Science Foundation (IIS-2145164, CCF-2212519), the DARPA Quantum Inspired Classical Computing (QUICC) program, and an Intel Rising Star Faculty Award. U.S. was financially supported out of the Saxonian State budget approved by the delegates of the Saxon State Parliament. S.K. acknowledges the HZO Si FE-FET device development work originated from the SRC funding received under the SRC GRC task 2825.001. The NSF, Grant #EEC-2123863, also supported part of this collaborative work. Any opinions, findings, and conclusions or recommendations expressed in this material are those of the author(s) and do not necessarily reflect the views of the NSF. The authors thank Roy H. Olsson at UPenn to allow them access to probe station in his lab for electrical measurements.

**AUTHOR CONTRIBUTIONS**

X.L., K.K., D.J., and P.C. conceived the idea. X.L. conceptualized and designed the FeFET-based ACAM circuits and in-memory search architecture. X.L. and K.K. performed machine learning simulations. X.L. and Y.H. performed electrical meaurements and SPICE simulations. P.J. performed microfabrication of FeFETs under supervision of S.K., U.S., and C.R.. X.L., K.K., D.J., and P.C. analyzed and interpreted the data. All authors contributed to writing of the manuscript.

**DECLARATION OF INTERESTS**

X.L., K.K., D.J., and P.C. have filed a provisional patent application based on the idea. The authors declare no other competing interests.

# Supplementary Information
# Analog Content-Addressable Memory from Complementary FeFETs

Xiwen Liu [a‡], Keshava Katti [a‡], Yunfei He [a], Paul Jacob [b], Claudia Richter [c], Uwe Schroeder [c], Santosh Kurinec [b], Pratik Chaudhari [a]*, Deep Jariwala [a]*.

[a] Electrical & Systems Engineering, University of Pennsylvania, Philadelphia, PA, USA.
[b] Electrical & Microelectronic Engineering, Rochester Institute of Technology, Rochester, NY, USA.
[c] Namlab/TU Dresden, Noethnitzer Strasse 64a 01187 Dresden, Germany.
*Corresponding authors: pratikac@seas.upenn.edu, dmj@seas.upenn.edu.
‡ Equal contribution.

**Measurement Configuration of the Single ACAM Cell**

The current-voltage characteristics of the ACAM cell were determined using a Keithley 4200A semiconductor characterization system. A single ACAM unit configuration, composed of one NMOS and one PMOS FeFET, is depicted in Fig. 1. Within a single ACAM cell, the two FeFETs are connected in parallel. The gates are both connected to the force terminal of the Keithley 4200A system, and the drain of the NMOS and PMOS FeFETs are attached to the sense terminal. Lastly, the source of the NMOS and PMOS FeFETs are grounded. The cabling setup utilizes tee adaptors to merge two wires for each terminal. A schematic of a single FeFET along with dimensions is shown in Figure 1b. A process flow for the FeFET fabrication is also provided below.

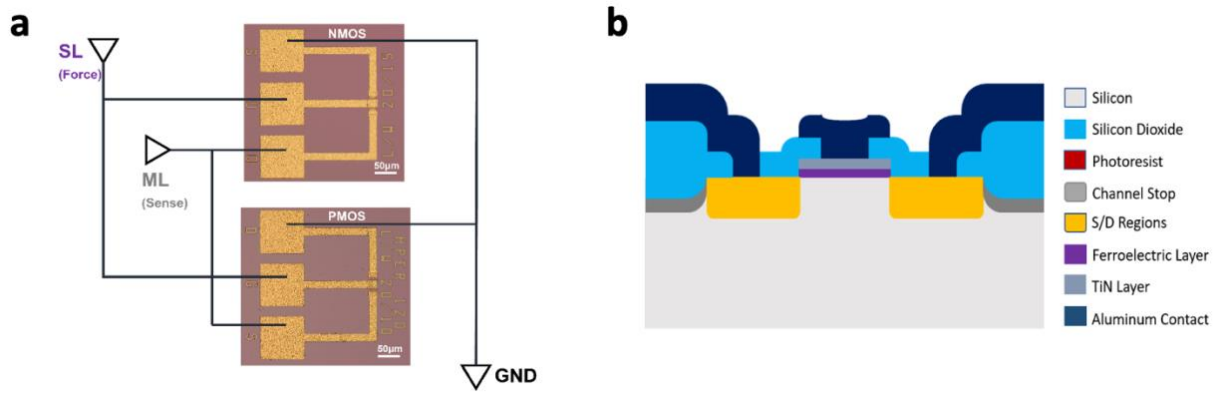

**Fig. 1. Electrical characterization setup for a single ACAM cell.** (**a**) This diagram depicts the configuration used to perform electrical characterization measurements on the single ACAM unit. (**b**) Schematic of single FeFET and its associated regions.

**FeFET Fabrication**

The highlights of the fabrication process are listed below.

- 6 inch (150 mm) wafer
- Photomask set → 5 level → active, source/ drain
- Contact cut, gate, metal
- Photomasks compatible with ASML i-line lithography stepper
- Channel stop implant
- LOCOS isolation
- S/D ion implantation using Varion 350D Implanter
- Gate layer ALD and RTA for ferroelectric

A full list of steps is also shown below. Acronyms used are standard Si CMOS micro-nanofabrication process acronyms.

1) RCA Clean
2) Zero Level Litho
3) Zero Level Etch
4) Resist Strip
5) RCA Clean
6) Pad Oxide Growth
7) LPCVD Nitride Deposition
8) Level 1 Litho (Active)
9) Nitride Etch
10) Channel Stop Implant (B11, 8e13)
11) Resist Strip
12) RCA Clean
13) Field Oxide Growth
14) Nitride Etch
15) Pad Oxide Etch
16) Kooi Oxide Growth
17) Level 2 Litho (S/D)
18) S/D Implant
19) Resist Strip
20) RCA Clean
21) S/D Anneal
22) Level 3 Litho (CC)
23) Kooi Oxide Etch
24) Resist Strip
25) RCA Clean
26) Ferro Deposition
27) Ferro Anneal
28) Level 4 Litho (Gate)
29) Ferro Etch
30) Resist Strip
31) BOE Etch
32) Al Deposition
33) Level 5 Litho (Metal)
34) Al Etch
35) Resist Strip
36) Freckle Etch
37) Al Sinter

**Circuit-Level Simulation for ACAM Arrays**

Transient simulations for a 64×64 FeFET-based ACAM array during a search operation are performed using SPICE (Fig. 2a). For each match line (ML), the 64 ACAM units have an individual search line ($SL_1 - SL_N$, where N = 64) and share a common match line. The ACAM units are preprogrammed such that a "match" state, marked by a high voltage on the ML, is maintained when the SL voltage ($V_{SL1} - V_{SLN}$, where N = 64) is in [0.4, 0.6] V. A match line sense amplifier (MLSA) is incorporated at the end of the ML to amplify the voltage difference during a mismatch. The MLSA, illustrated in Fig. 2b, is designed as a single-stage amplifier with a current mirror differential pair using ~45 nm technology, in which parasitic parameters of 45 nm technology node are used for simulation. The transient operation's output voltage for the ACAM array, post-amplification, is displayed in Fig. 2c. For the worst case, to simulate a single-bit mismatch, a mismatch voltage is applied at $SL_1$ ($V_{SL1}$ = 0.7 V at t = 1 ns). The output voltage at ML discharges to a low level, indicating a "mismatch" state with a delay time of 78.3 ps.

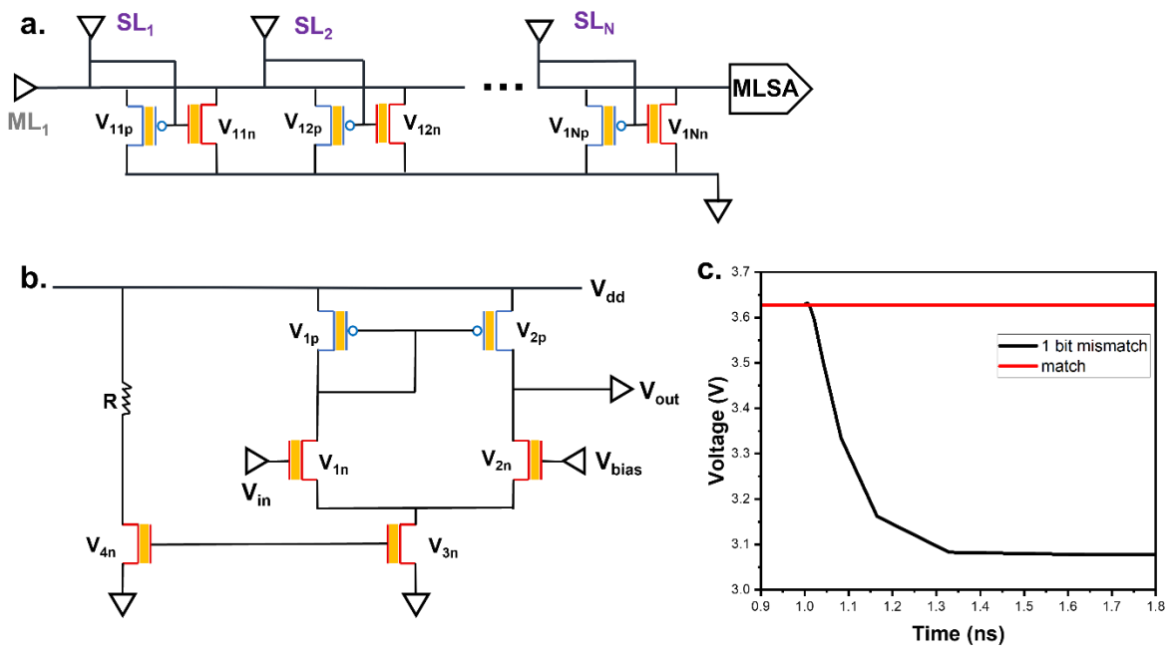

**Fig. 2. SPICE simulation for a 64×64 ACAM array and the related schematic diagrams.** (**a**) Circuit schematic illustrating the simulation of 1×64 ACAM array during search operation response to mismatch. (**b**) Schematic representation of match line sense amplifier (MLSA) design, which features a differential pair and current mirror using ~45 nm technology. (**c**) Transient response of output voltage under single-bit mismatch scenario, demonstrating transition of match line ($ML_1$) from "match" state to "mismatch" state with delay time of 78.3 ps.

**Bit-Density Analysis of ACAM Design**

We will now discuss how ACAM could facilitate strong bit densities and how a single ACAM cell could represent equivalent bits within TCAM. To accomplish superior bit density, a single ACAM cell should feature more non-overlapping matching windows and narrower windows, as illustrated in Fig. 3a(1)-(2). However, these two characteristics set up a trade-off between bit density and sense margin. As bit density increases, the width of the matching windows decreases, leading to a smaller sense margin between match and mismatch states, given the non-infinite slope of the transistor in the sub-threshold region, as depicted in Fig. 3a(3)-(4). This phenomenon is known as the thermionic limit, and it equates to approximately 60 mV/dec at room temperature (300 K). The relationship between sense margin and bit density can be modeled quantitatively. The logarithm of sense margin linearly decreases with bit density, as shown in Fig. 3b. Based on SPICE simulations, we find that a 3-bit density of ACAM could be achieved for a targeted sense margin of 1,000, while a 4-bit density of ACAM could be realized for a targeted sense margin of 100.

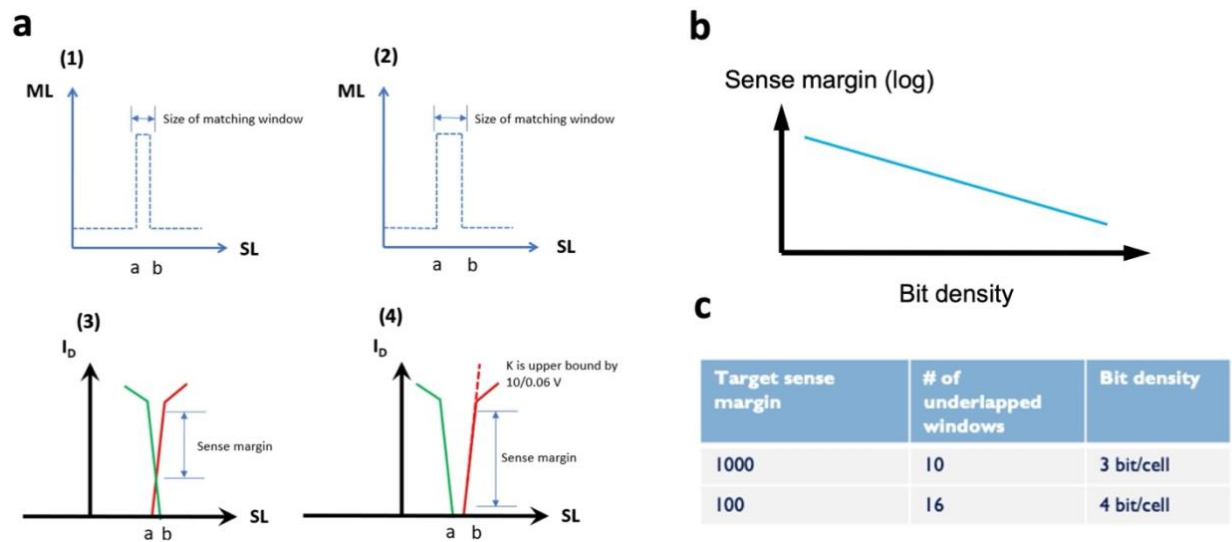

**Fig. 3. Bit-density analysis of ACAM design.** (**a**) Attaining superior bit density necessitates inclusion of more non-overlapping matching windows within single ACAM cell, coupled with reduction of matching window width. However, this approach precipitates a trade-off between bit density and sense margin. With increase in bit density, width of matching windows diminishes, subsequently leading to reduction in sense margin between match and mismatch states. (**b**) Quantitatively, relationship between sense margin and bit density can be expressed as follows. As bit density escalates, logarithm of sense margin decreases linearly. (**c**) According to SPICE simulations, varying levels of bit density can be achieved in ACAM, each corresponding to different targeted sense margins.

**Software Implementation of Matching Networks for Few-Shot Learning**

For the software-based simulations, the CPU is the Intel Xeon Platinum (14 nm node), while the GPU is the NVIDIA Tesla T4 (12 nm node). In matching networks, data are partitioned into *n*-way, *k*-shot, *q*-query episodes. Such an episode contains *n* classes, from which there are *k* samples. These $n \cdot k$ samples make up the support set. The *q* query samples make up the query set, where the query samples also belong to one of the *n* classes in the support set. Together, the support set and query set form an episode. In this work, *q* will always be set to 1 and will hence be omitted when referring to an episode.

During training, the entire training dataset is partitioned into episodes, akin to mini-batches. The support set is then passed through a 4-layer convolutional neural network (CNN) that is known as the backbone, yielding 64-dimensional embeddings. Note that the backbone and all episodic data are transferred to the GPU. Each layer in the backbone consists of a 3×3 kernel with 64 filters, batch normalization, ReLU non-linearity, and a 2×2 max-pooling. The backbone outputs a 64-dimensional embedding for each support sample. These support embeddings are then stored in an array of size $n \cdot k \times 64$. The query sample is also passed through the backbone, stored as a single 64-dimensional array. Once the support and query embeddings are stored in the respective arrays within software, what remains is for the cosine similarity $S_{\text{cosine}}(\mathbf{s}_i, \mathbf{q}) = \mathbf{s}_i \cdot \mathbf{q} \,/\, \|\mathbf{s}_i\|\|\mathbf{q}\|$ to be computed for all $i = 1, \ldots, n \cdot k$, where $\mathbf{s}_i$ is the *i*th support embedding and $\mathbf{q}$ is the query embedding. After which, $\operatorname{argmax}_{i=1,\ldots,n \cdot k}\{S_{\text{cosine}}(\mathbf{s}_i, \mathbf{q})\}$ returns the index of the support sample that is the most similar to the query sample, and its label is predicted as that of the query. Put simply, the query embedding is compared with each of the support embedding, with the label of the closest match given to the query sample.

Error is evaluated using cross-entropy loss and back-propagated through the backbone for every episode in the training process. Inference involves receiving episodic test data for which this same process is used to classify the query sample. The inference process can be summarized as a (cosine) similarity search and is demonstrated on both CPU and GPU.

**Kernel Definition and ACAM Implementation of Kernel Regression**

A Mercer kernel is symmetric, continuous function $K(\mathbf{x},\mathbf{x}'): \mathcal{X} \times \mathcal{X} \to \mathbb{R}$. In this work, the kernel function is defined on the Euclidean instance space $\mathcal{X} = \mathbb{R}^d$. The output of a Mercer kernel captures the similarity between $d$-dimensional vectors $\mathbf{x}$ and $\mathbf{x}'$. The training data can be written as $\mathcal{D}_{\text{Tr}} = \{(x_i, y_i)\}_{i=1}^m$. A kernel regression model requires computing the pairwise similarity between each training data point and storing it in a matrix $\mathbf{K}[i,j] = K(\mathbf{x}_i, \mathbf{x}_j)$, where $\mathbf{K} \in \mathbb{R}^{m \times m}$ is called a Gram matrix so long as it is positive semi-definite. This work is interested in specifically a Gaussian kernel, also known as the radial basis function (RBF) kernel, $K(\mathbf{x},\mathbf{x}') = \exp(-\|\mathbf{x}-\mathbf{x}'\|_2^2 / 2\gamma^2)$. The training process is then described by the following equation, $\hat{\boldsymbol{\alpha}} = (\mathbf{K} + \lambda m \mathbf{I}_m)^{-1}\mathbf{y}$, where $\mathbf{K}$ is the Gram matrix, $\lambda$ is a regularizer parameter, $m$ is the number of training data points, $\mathbf{I}_m$ is the $m \times m$-dimensional identity matrix, and $\mathbf{y}$ is the vector of training data labels. The regularizer parameter $\lambda$ tunes how much importance is placed upon generating a "simple" solution $\hat{\boldsymbol{\alpha}}$. A simple solution can be thought of as one where weights $\hat{\boldsymbol{\alpha}}$ remain close to the origin. A larger $\lambda$ yields a simpler solution but typically at the cost of more bias. As a result, it is standard to perform cross-validation to determine the best $\lambda$ parameter.

Following the expanded definition of a Mercer kernel, Gram matrix, and regularization, the remainder of this section will expand on the ACAM implementation of kernel regression. First, recall that a constant-time procedure is described to obtain the surrogate Gaussian kernel. Fig. 4a visualizes this surrogate Gaussian kernel (red) alongside a conventional Gaussian kernel (purple). Notice that both functions are qualitatively similar. Second, in the runtime benchmark for kernel regression experiments, the reason that CPU and GPU are combined into a single category is that, on average, both had a very comparable runtime for kernel regression. As seen in Fig. 2d, the vast majority of runtime is involved in computing the kernel, which is a 2-loop iterative function that calls exponential and norm functions $O(m^2)$ times. A 64×64 ACAM array (45 nm node) was used for benchmarking. We will note again that the CPU is the Intel Xeon Platinum (14 nm node), while the GPU is the NVIDIA Tesla T4 (12 nm node). In the case of the GPU, its advantages could not be realized in such a small-scale computation. Only when the dimension of the data exceeded 100,000 did the GPU begin to perform better than the CPU, but even then, the GPU remained about 3 orders of magnitude slower than ACAM. Since data ranging from dimensions 1 – 100,000 are likely to be the most common in real-world settings, the regime where the CPU and GPU perform approximately equivalently on kernel regression inference, it seemed appropriate to group the two into a single column.

Lastly, in order to justify the claim that a dataset can be scaled in order to ensure that $\gamma = 0.1$ V is the optimal parameter, two experiments were executed, whose results are summarized in Figs. 4b-c. Particularly, scaling $x$ data by $k$ scales $\gamma$ to $k\gamma$. In Fig. 4b, the blue data is sampled from $f(x) = \sin(10x)$ and predicted using ACAM-based kernel regression with optimal parameter $\gamma = 0.05$ V. The $x$-values of the blue data are then scaled from the range [0.0, 1.0] V to [0.0, 2.0] V, shown by green scattered points, and predicted using ACAM-based kernel regression with optimal parameter $\gamma = 0.1$ V. Both the predictions on the blue data and those on the green data yielded identical mean squared error (MSE). As a result, Fig. 4b represents one instance of how $x$ data with optimal parameter $\gamma = 0.05$ V can be scaled by $k = 2$ to exhibit the desired $\gamma = k(0.05) = 0.1$ V required for ACAM. Fig. 4c provides further confirmation that scaling $x$ data by $k$ scales $\gamma$ to $k\gamma$. Data from $f(x) = \sin(20x)$ with optimal parameter $\gamma = 0.025$ V (yellow), $f(x) = \sin(10x)$ with optimal parameter $\gamma = 0.05$ V (purple), $f(x) = \sin(5x)$ with optimal parameter $\gamma = 0.1$ V (red), $f(x) = \sin(2.5x)$ with optimal parameter $\gamma = 0.2$ V (green), and $f(x) = \sin(1.125x)$ with optimal parameter $\gamma = 0.4$ V (blue) are all used for prediction. MSE is collected for all 5 cases and plotted on the x-axis of Fig. 4c. Then, data from $f(x) = \sin(20x)$ with $\gamma = 0.1$ V scaled to [0.0, 4.0] V (yellow), $f(x) = \sin(10x)$ scaled to [0.0, 2.0] V with $\gamma = 0.05$ V (purple), function $f(x) = \sin(5x)$ unscaled (i.e., [0.0, 1.0] V) with $\gamma = 0.1$ V (red), $f(x) = \sin(2.5x)$ scaled to [0.0, 0.5] V with $\gamma = 0.2$ V (green), and $f(x) = \sin(1.125x)$ with $\gamma = 0.1$ V scaled to [0.0, 0.25] V (blue) are all used for prediction. MSE is collected for all 5 cases and plotted on y-axis. The result the linear plot $y = x$, which implies that, for each of the 5 cases where the data was scaled, $\gamma = 0.1$ V is the optimal parameter.

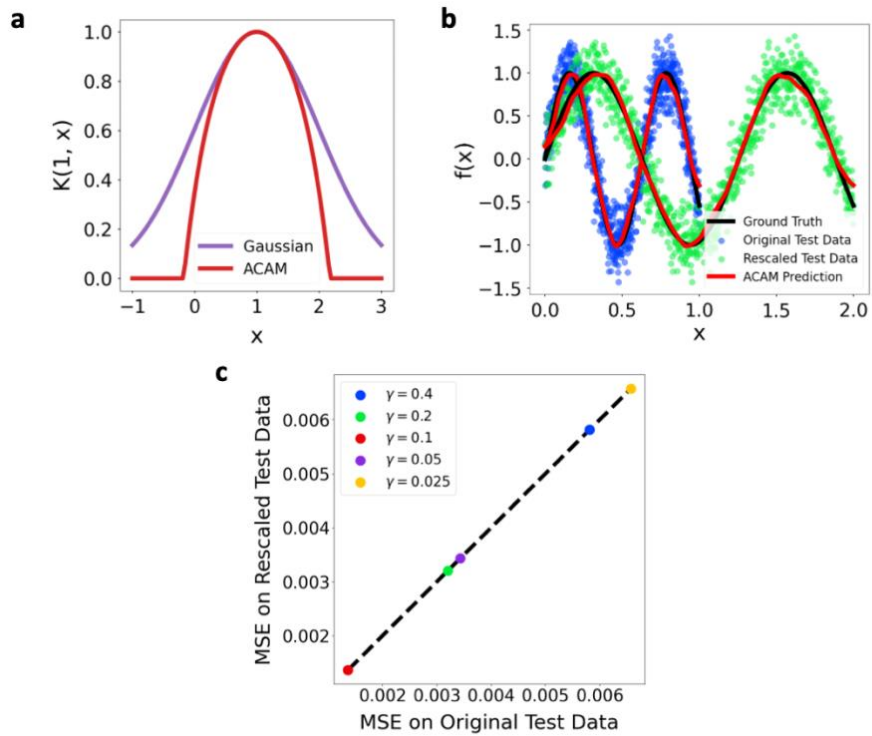

**Fig. 4. Surrogate Gaussian kernel and mapping dataset to $\gamma = 0.1$.** (**a**) Gaussian kernel $K(\mathbf{x}, \mathbf{x'}) = \exp(-\|\mathbf{x} - \mathbf{x'}\|_2^2 / 2\gamma^2)$ (purple) and surrogate Gaussian kernel $K^{ACAM}(\mathbf{x}, \mathbf{x'}) = max\{0, 2 - \exp(|\mathbf{x} - \mathbf{x'}|^2 / 2\gamma^2)\}$ (red) overlayed on same plot, where $\mathbf{x} = \mathbf{1}$ and $\mathcal{X} = \mathbb{R}$. (**b**) Blue data sampled from $f(x) = \sin(10x)$ and predicted with optimal parameter $\gamma = 0.05$ V. The x-values of blue data then scaled from range [0.0, 1.0] V to [0.0, 2.0] V, shown by green scattered points, and predicted with optimal parameter $\gamma = 0.1$ V. Both the predictions on blue data and green data yielded identical mean squared error (MSE). (**c**) x-axis shows MSE for predictions on data generated from $f(x) = \sin(20x)$, $f(x) = \sin(10x)$, $f(x) = \sin(5x)$, $f(x) = \sin(2.5x)$, and $f(x) = \sin(1.125x)$, with optimal parameters $\gamma = 0.025$ V, $\gamma = 0.05$ V, $\gamma = 0.1$ V, $\gamma = 0.2$ V, and $\gamma = 0.4$ V, respectfully. y-axis shows MSE for predictions on data generated from same 5 cases but with data scaled to [0.0, 4.0] V, [0.0, 2.0] V, [0.0, 1.0] V, [0.0, 0.5] V, and [0.0, 0.25] V, respectfully, all of which have optimal parameter $\gamma = 0.1$ V. Since result is linear plot $y = x$, it shows that scaling x data by $k$ indeed scales $\gamma$ to $k\gamma$.